\documentclass[preprintnumbers,amsmath,amssymb,preprint, nofootinbib]{revtex4}
\usepackage{graphicx}% Include figure files
\usepackage{amsmath,amssymb,graphics,epsfig}
\usepackage{dcolumn}% Align table columns on decimal point
\usepackage{epsf}
\usepackage{epstopdf}
\usepackage {amssymb}
\newcommand{\nc}{\newcommand}
\nc{\ba}{\begin{eqnarray}}
\nc{\ea}{\end{eqnarray}}
\newcommand\be{\begin{equation}}
\newcommand\ee{\end{equation}}

\newcommand{\calP}{{\cal P}}

\newcommand{\zetad}{{\zeta_{\rm dec}}}
\newcommand\mPl{{M_{\rm Pl}}}

\begin{document}

\title{Modulated curvaton decay}
\author{Hooshyar Assadullahi$^{1, 2}$}
\email{hooshyar.assadullahi-AT-port.ac.uk}
\author{Hassan Firouzjahi$^{3}$}
\email{firouz-AT-mail.ipm.ir}
\author{Mohammad Hossein Namjoo$^{4, 5}$}
\email{mh.namjoo-AT-mail.ipm.ir}
\author{David Wands$^{1}$}
\email{david.wands-AT-port.ac.uk}
\affiliation{$^1$ Institute of Cosmology and Gravitation, University of Portsmouth, Dennis Sciama Building, Burnaby Road, Portsmouth PO1 3FX, United Kingdom}
\affiliation{$^2$ School of Earth and Environmental Sciences, University of Portsmouth, Burnaby Building, Burnaby Road, Portsmouth PO1 3QL, United Kingdom}
\affiliation{$^3$School of Astronomy, Institute for Research in
Fundamental Sciences (IPM),
P.~O.~Box 19395-5531,
Tehran, Iran}
\affiliation{$^4$Yukawa Institute for theoretical Physics,
 Kyoto University, Kyoto 606-8502, Japan}
\affiliation{$^5$School of Physics, Institute for Research in
Fundamental Sciences (IPM),
P.~O.~Box 19395-5531,
Tehran, Iran}

\begin{abstract}
\vspace{0.3cm}
We study primordial density perturbations generated by the late decay of a curvaton field whose decay rate may be modulated by the local value of another isocurvature field, analogous to models of modulated reheating at the end of inflation. We calculate the primordial density perturbation and its local-type non-Gaussianity using the sudden-decay approximation for the curvaton field, recovering standard curvaton and modulated reheating results as limiting cases. We verify the Suyama-Yamaguchi inequality between bispectrum and trispectrum parameters for the primordial density field generated by multiple field fluctuations, and find conditions for the bound to be saturated.
\vspace{0.3cm}
%Keywords :  Gravity Waves, Cosmic Strings
\end{abstract}

\date\today

\preprint{IPM/A-2012/019 }

\maketitle

\section{Introduction}

Inflation in the very early universe provides a classical cosmology that drives the universe towards spatial flatness and homogeneity. It also provides an origin for primordial density perturbations through the quantum fluctuations of light scalar fields, stretched by the inflationary expansion to super-Hubble scales. Originally structure was assumed to originate from fluctuations in the inflaton field driving inflation, but more recently it has been realised that there are many possible mechanisms through which scalar field fluctuations could generate the observed primordial density perturbations \cite{Bassett:2005xm}. Examples incude the late decay (some time after inflation has ended) of a curvaton field \cite{Linde:1996gt,Enqvist:2001zp,Lyth:2001nq,Moroi:2001ct}, modulated reheating at the end of inflation, where the rate of inflaton decay is modulated by the local vacuum expectation value (VEV) of an isocurvature field \cite{Kofman:2003nx,Dvali:2003em}, or an inhomogeneous end of inflation \cite{Bernardeau:2004zz,Lyth:2005qk}. 
All of these mechanisms for the origin of the primordial density perturbations may be distinguished from adiabatic perturbations in the inflaton field driving inflation as they  introduce local-type primordial non-Gaussianity \cite{Wands:2010af}, most simply characterised by the non-linearity parameter, $f_{NL}$ \cite{Komatsu:2001rj}, which typically becomes large when the transfer efficiency becomes small \cite{Alabidi:2010ba}.

In this paper we go beyond the simplest curvaton models to include the possible modulation of the curvaton decay by a modulator field, $\chi$, in analogy with modulated reheating at the end of inflation.
In modulated reheating scenarios the rate of decay of a massive field $\sigma$ is assumed to be dependent on the VEV of a modulator field, $\chi$. Consider a Lagrangian which describes the decay of the $\sigma$-field oscillations into $\psi$-particles
\be
\label{L}
{\cal L} =
% -\frac12 m_\sigma^2 \sigma^2
 - V(\sigma)
 - \lambda(\chi)\sigma\bar \psi \psi - U(\chi) \,.
\ee
If the field, $\sigma$, is displaced from its minimum during inflation, then the field oscillates once $H<m_\sigma$ where $m_\sigma^2=V''(\sigma)$ at its minimum. Note that the potential, $V(\sigma)$, may deviate from a simple quadratic potential, but we will assume it can be described by a simple quadratic potential once it begins to oscillate about its minimum, with effective mass $m_\sigma$.
If the effective mass of the $\chi$ field remains small while the $\sigma$ field oscillates, $|U''(\chi)|<H^2<m_\sigma^2$, then the decay rate of massive $\sigma$-particles into light $\psi$-particles is given by $\Gamma\propto\lambda^2$ which is assumed to be a function of the local VEV of $\chi$. For example, if the coupling $\lambda$ is a linear function of $\chi$ then the decay rate is a quadratic function $\Gamma\propto\chi^2$.

Any light field during inflation acquires an almost scale-invariant distribution of perturbations due to small-scale vacuum fluctuations being stretched up and frozen-in on super-Hubble scales. In particular, we will assume that light fields acquire a Gaussian distribution of perturbations at Hubble-exit ($k=(aH)_*$) with power spectrum
\be
{\cal P}_* = \left( \frac{H_*}{2\pi}\right)^2 \,.
\ee
Local variations in the $\chi$ field change the local rate of reheating and hence the primordial density perturbations, which can be represented by $\zeta$, the metric perturbation on uniform-density hypersurfaces \cite{Malik:2008im} in the primordial radiation-dominated era, some time after inflation has ended.

In the original modulated reheating scenario $\sigma$ is the inflaton whose oscillations dominate the energy density of the universe immediately after inflation comes to an end. In this work we will consider the case where $\sigma$ is a curvaton field whose energy density is sub-dominant during inflation. We assume its effective mass is small compared to the Hubble scale during inflation and hence it also acquires an almost scale-invariant spectrum of fluctuations. The curvaton mass becomes larger than the Hubble scale either immediately at the end of inflation or some time after, when $H=m_\sigma$ and the curvaton field eventually decays. Fluctuations in the local VEV of the curvaton during inflation lead to perturbations in the amplitude of oscillations and the density of curvaton particles when the field oscillates after inflation. These density perturbations are transferred to the radiation when the curvaton decays.

If both the modulator field and the curvaton field are light during inflation then their vacuum fluctuations can generate a primordial density perturbation, in addition to any adiabatic density perturbations produced from adiabatic fluctuations in the inflaton field, $\phi$, during inflation.

The outline of this paper is as follows. In section \ref{sect:suddendecay} we describe how we can estimate the primordial density perturbations produced by the modulated curvaton decay in terms of the inhomogeneous densities and metric on an instantaneous decay hypersurface, generalising previous results for non-linear perturbations from curvaton decay with an unmodulated decay rate \cite{Sasaki:2006kq}. We calculate observable quantities including the primordial power spectrum, tensor-scalar ratio, bispectrum and trispectrum. We show that the Suyama-Yamaguchi inequality between the tree-level bispectrum and trispectrum holds and consider the condition for this inequality to be saturated. We present our conclusions and discussion in section~\ref{sect:discuss}.

\section{Sudden-decay approximation}
\label{sect:suddendecay}

We will work in the sudden-decay approximation where the curvaton decay is modelled as an instantaneous transfer of energy from the curvaton field oscillations, $\rho_\sigma$, into radiation, $\rho_\gamma$. This has been shown to be a good approximation to the full numerical results in the usual curvaton scenario \cite{Sasaki:2006kq}.

If the curvaton decay rate is a constant, $\Gamma=\bar\Gamma$, then the decay hypersurface, $H_{\rm dec}=\Gamma$, corresponds to a uniform-density hypersurface (on super-Hubble scales at the decay time) with
\be
\bar\rho_{\rm dec} = 3\mPl^2 \bar\Gamma^2 \,.
\ee
In the presence of fluctuations in the local VEV of $\chi$, the modulator field, and therefore local fluctuations of the decay rate, we have
\be
\label{rho-decay}
\rho_{\rm dec} = 3\mPl^2 \Gamma^2(\chi) \,.
\ee

We will allow for the existence of inhomogeneous perturbations of the radiation density, $\zeta_\gamma$,  after inflation but before the curvaton decays, due to adiabatic inflaton field fluctuations during inflation, $\zeta_\gamma=\zeta_\phi=-H \delta\phi/\dot\phi$.
Before the curvaton decay, the curvature perturbation on uniform-radiation-density hypersurfaces, $\zeta_\gamma$, and uniform-curvaton-density hypersurfaces, $\zeta_\sigma$, are independently conserved \cite{Lyth:2003im}.  After the decay, assuming the decay products are relativistic, the curvature perturbation on uniform-total-density hypersurfaces, $\zeta$, is conserved.
On the decay hypersurface itself, we therefore have \cite{Lyth:2004gb,Sasaki:2006kq} 
\ba
\label{rho-before}
%\rho  \simeq  \rho_\sigma  + \rho_\gamma
%\qquad , \qquad
 \rho_{\sigma,{\rm dec}} = \bar \rho_{\sigma,{\rm dec}} e^{3(\zeta_\sigma-\zetad)} 
\, , \qquad 
\rho_{\gamma,{\rm dec}} = \bar \rho_{\gamma,{\rm dec}} e^{4(\zeta_\gamma-\zetad)}  
\,, \qquad 
\label{rho-after}
\rho_{\rm dec} = \bar \rho_{\rm dec} e^{4(\zeta-\zetad)} \,,
\ea
where $\zetad$ is the curvature perturbation ($\delta N$ from a flat hypersurface) on the decay hypersurface. Note that we neglect the energy density of the modulator field $\chi$ throughout.

Matching
% \eqref{rho-after} and \eqref{rho-before} 
$\rho_\gamma+\rho_\sigma=\rho$ 
at the decay time
%, and using \eqref{rho-decay}, 
yields
\ba
\label{full}
 (1-\Omega_{\sigma,{\rm dec}}) e^{4 (\zeta_\gamma-\zetad)} + \Omega_{\sigma,{\rm dec}} e^{3(\zeta_\sigma-\zetad)} = e^{4(\zeta-\zetad)} \,,
\ea
where
\ba
\Omega_\sigma \equiv \dfrac{\bar \rho_\sigma}{\bar \rho} \,.
\ea

\subsection{Linear perturbations}

At linear order, one can expand the above relation (\ref{full}) to give
\ba
 - \zetad \simeq \dfrac{1}{\Omega_{\sigma,{\rm dec}}} \left[3 \Omega_{\sigma,{\rm dec}} \zeta_\sigma + 4 (1- \Omega_{\sigma,{\rm dec}}) \zeta_\gamma -4 \zeta \right]
\ea
Using the above relation to eliminate $\zetad$ in \eqref{rho-after} and also using \eqref{rho-decay}, results in
\ba
 \label{eq:zeta}
\zeta \simeq \zeta_\gamma  - f \, \dfrac{\delta \Gamma}{6 \Gamma} + f ( \zeta_\sigma - \zeta_\gamma ) \,,
\ea
where we have defined the transfer parameter
\ba
 \label{fdef}
f  \equiv \dfrac{3 \Omega_{\sigma,{\rm dec}}}{4 - \Omega_{\sigma,{\rm dec}}}  \,.
\ea

The change in the curvature perturbation, $\zeta-\zeta_\gamma$ in Eq.~(\ref{eq:zeta}), due to the modulated curvaton decay is seen to arise from the relative entropy perturbation between the curvaton density and the radiation density, $\zeta_\sigma-\zeta_\gamma$, and the perturbed decay rate, $\delta\Gamma\simeq\Gamma'\delta\chi$. 
In the case of homogeneous curvaton decay rate, $\delta\Gamma=0$, we reproduce the standard curvaton result. The inhomogeneous decay rate adds an extra term in the primordial curvature perturbation~(\ref{eq:zeta}) proportional to $f$.

Note that since $\chi$ remains overdamped throughout the decay, we assume that we can neglect its background evolution and hence its perturbation on the decay hypersurface correspond directly to its perturbation on spatially flat hypersurfaces during inflation. The curvaton, $\sigma$, is overdamped during inflation, but starts to oscillate when $H=m_\sigma$. We assume the curvaton density is still negligible at this time, so the curvaton field fluctuations on this surface correspond to curvaton density perturbations on a uniform radiation-density hypersurface
\be
\rho_\sigma = \bar\rho_\sigma e^{3(\zeta_\sigma-\zeta_\gamma)} \,.
\ee
Allowing for possible evolution of the curvaton field from the end of inflation up until the point at which it starts oscillating we write
\be
\rho_\sigma = \frac12 m_\sigma^2 g^2(\bar\sigma+\delta\sigma) \,.
\ee
At linear order we thus have
\be
S_\sigma \equiv 3(\zeta_\sigma - \zeta_\gamma ) = 2\frac{g'\delta\sigma}{g} \,.
\ee
where $\bar\sigma+\delta\sigma$ describes the local curvaton VEV at the end of inflation.

\subsection{Non-linear perturbations}

Expanding \eqref{rho-decay} and \eqref{rho-after} to second order yields
\ba
\zeta -\zetad = \dfrac{1}{2} \ln \left(\dfrac{\Gamma}{\bar \Gamma} \right) \simeq \dfrac{\Gamma'}{2\Gamma} \delta \chi + \dfrac{1}{4} \left( \dfrac{\Gamma''}{\Gamma} -\dfrac{\Gamma'^2}{\Gamma^2}\right) \delta \chi^2
\ea
Eliminating $\zetad$ in \eqref{full} and solving for $\zeta$ order by order yields
% \ba
% \zeta &=& - f \, \dfrac{ \Gamma'}{6 \Gamma} \delta \chi + f \zeta_\sigma + (1-f) \zeta_\gamma + 
% \dfrac{f(1-f)(3+f)}{2} (\zeta_\gamma - \zeta_\sigma)^2 + \dfrac{\Gamma' \, f (1-f) (3+f)}{6 \Gamma }  
% (\zeta_\gamma - \zeta_\sigma) \delta \chi  
% \\ \nonumber
% &+& \dfrac{f}{36} \left( -3 \dfrac{\Gamma''}{\Gamma} +\dfrac{\Gamma'^2}{2 \Gamma^2} (9-f(f+2))\right) % \delta \chi^2
% \ea
% which results in
\ba
 \label{zeta2}
\zeta &\simeq& \zeta_\gamma + \dfrac{f}{3} S_\sigma  - f \, \dfrac{ \Gamma'}{6 \Gamma}\delta \chi + \dfrac{f(1-f)(3+f)}{18} S_\sigma^2 - \dfrac{\Gamma' \, f (1-f) (3+f)}{18 \Gamma }  S_\sigma \delta \chi  
\\ \nonumber
&+& \dfrac{f}{36} \left( -3 \dfrac{\Gamma''}{\Gamma} +\dfrac{\Gamma'^2}{2 \Gamma^2} (9-f(f+2))\right) \delta \chi^2
\ea
where we define the entropy isocurvature perturbation
\ba
S_\sigma \equiv 3 (\zeta_\sigma - \zeta_\gamma) \simeq 2 \dfrac{g'}{g} \delta \sigma 
-\left( \dfrac{g'}{g}\right)^2  \left[ 1 - \dfrac{g''g}{g^{\prime2}} \right] \delta \sigma^2 \,.
\ea

%Finally then we can express the second-order primordial density perturbation in terms of the field fluctuations during inflation
%\ba
%\zeta &\simeq& \zeta_\gamma + \dfrac{2f}{3} \frac{g'}{g} \delta\sigma - f \, \dfrac{ \Gamma'}{6 \Gamma}\delta \chi + \dfrac{f(1-f)(3+f)}{18} S_\sigma^2 - \dfrac{\Gamma' \, f (1-f) (3+f)}{18 \Gamma }  S_\sigma \delta \chi   \\ \nonumber
%&+& \dfrac{f}{36} \left( -3 \dfrac{\Gamma''}{\Gamma} +\dfrac{\Gamma'^2}{2 \Gamma^2} (9-f(f+2))\right) \delta \chi^2 \ea

\subsection{Observables}

When calculating observables, such as the power spectrum and higher-order correlators of the primordial density perturbations, it will be convenient to express the non-linear perturbation $\zeta$ in terms of the perturbed logarithmic expansion, $N=\int H dt$, from an initial spatially flat hypersurface where the scalar field perturbations originate during inflation and a final uniform-density hypersurface during the subsequent radiation-dominated era \cite{Lyth:2004gb,Lyth:2005fi}
we note that the total curvature perturbation can be rewritten by
\ba
 \label{zetaN}
\zeta = N_a \delta \phi^a + \dfrac{1}{2} N_{ab} \delta \phi^a \delta \phi^b 
+ \dfrac{1}{6} N_{abc} \delta \phi^a \delta \phi^b \delta \phi^c +...
\ea
where $\delta \phi^a$ are the three Gaussian fields $\delta \phi$, $\delta \chi$ and $\delta \sigma$ , $N_a = N^a \equiv \partial N/\partial \phi_a$ and 
$N_{ab}= N^{ab} \equiv \partial^2 N/\partial \phi_a \phi_b$. 

Comparing Eq.~(\ref{zetaN}) with (\ref{zeta2}) we identify
\ba
\label{Na}
N_\phi = \dfrac{1}{\sqrt{2 \mPl^2 \epsilon_\phi}} \qquad , \qquad 
N_\sigma = \frac{2fg'}{3g} 
\qquad , \qquad 
N_\chi = - \frac{f\Gamma'}{6\Gamma}
% = - \dfrac{1}{\sqrt{2 \mPl^2 \epsilon_\chi}} 
% N_\sigma= \dfrac{1}{\sqrt{2 \mPl^2 \epsilon_\sigma}} 
\ea
\ba
\label{Nsigchi}
N_{\chi \sigma} &=& - \dfrac{f (1-f) (3+f)\Gamma' g'}{9\Gamma g} 
\\
\label{Nsigsig}
N_{\sigma \sigma} &=& \dfrac{2f}{3} \left[ 1 + \dfrac{g''g}{g^{\prime2}} - \frac{4f}{3} - \frac{2f^2}{3} \right] \left( \dfrac{g'}{g}\right)^2 
\\
\label{Nchichi}
N_{\chi \chi} &=& \dfrac{f}{36} \left[ 9-6 \dfrac{\Gamma''\Gamma}{\Gamma^{\prime2}}-2f-f^2 \right] \left( \dfrac{\Gamma'}{\Gamma}\right)^2 \,.
\ea
in which
\be
\epsilon_\phi \equiv - \left(\dfrac{\dot H}{H^2} \right)_*
\, .
\ee
Note that higher-derivatives with respect to the inflaton field $\phi$ can be neglected during slow-roll inflation.

{}From Eqs.(\ref{Na}-\ref{Nsigchi}) we identify
\ba
\label{fsig}
f_\sigma \equiv \frac{\partial f}{\partial \sigma}=  \frac{2f(1-f)(3+f)}{3} \frac{g'}{g} \,,\\
\label{fchi}
f_\chi  \equiv \frac{\partial f}{\partial \chi}= - \frac{f(1-f)(3+f)}{6} \frac{\Gamma'}{\Gamma}  \,,
\ea
This will allow us to calculate all higher derivatives of $N$ starting from Eqs.~(\ref{Na}-\ref{Nsigchi}).  One can verify this is consistent with Eqs.~(\ref{Nsigsig}-\ref{Nchichi}). 

The power spectrum is given, at leading order, by
\ba
\label{Pzeta}
\calP_\zeta = \calP_{\zeta_\phi} + \dfrac{f^2}{9} \calP_{S_\sigma} + f^2 \left(\dfrac{\Gamma'}{6 \Gamma} \right)^2 \calP_\chi
 =
 \dfrac{1}{2 \mPl^2} \left(\dfrac{1}{\epsilon_\phi} + \dfrac{1}{\epsilon_\chi} + \dfrac{1}{\epsilon_\sigma} \right) \left(\dfrac{H}{2\pi} \right)^2_*
\ea
in which, in analogy with the inflaton contribution, we have defined \cite{Fonseca:2012cj}
\ba
\epsilon_\sigma \equiv \dfrac{9}{8} \left(\dfrac{g}{f g' \mPl} \right)^2 \qquad , \qquad 
\epsilon_\chi \equiv  18 \left(\dfrac{  \Gamma}{f \Gamma' \mPl} \right)^2 
\ea
The relative contribution of each field to the power spectrum (\ref{Pzeta}) is given by the weights $w_a$ defined via ${\calP}_{\zeta_a} = w_a \calP_\zeta$ in which 
\be
 \label{wa}
w_a \equiv 
 \frac{N_a^2}{N_\phi^2+N_\sigma^2+N_\chi^2} 
=
\frac{\epsilon_a^{-1}}{\epsilon_\chi^{-1} + \epsilon_\phi^{-1}+\epsilon_\sigma^{-1}}
 \,.
\ee
Note that $ w_\phi + w_\sigma + w_\chi= 1$.

The spectral index is then
\ba
n_s-1 = w_\chi (n_{\zeta_\chi}-1 ) + w_\sigma (n_{\zeta_\sigma}-1 ) + (1-w_\chi-w_\sigma) (n_{\zeta_\phi}-1 ) 
\ea
% In which we defined
% \ba
% w_\sigma = \dfrac{ \epsilon_\chi \epsilon_\phi}{\epsilon_\chi + \epsilon_\phi+\epsilon_\sigma} \qquad , % \qquad 
% w_\chi = \dfrac{ \epsilon_\sigma \epsilon_\phi}{\epsilon_\chi + \epsilon_\phi+\epsilon_\sigma} 
% \ea
The tensor to scalar ratio is also given by
\ba
r = 16 \epsilon_\chi w_\chi = 16 \epsilon_\sigma w_\sigma = 16 (1-w_\sigma - w_\chi ) \epsilon_\phi \,.
\ea

The local-type primordial bispectrum is characterised at leading order in term of $f_{NL}$ \cite{Lyth:2005fi}
\ba
\dfrac{6}{5} f_{NL} = \dfrac{N_a N_b N^{ab}}{(N_a N^a)^2} 
\ea
which we can write as 
\be
\label{fNL}
f_{NL} = w_\sigma^2 f_{NL}^\sigma + 2 w_\chi w_\sigma f_{NL}^{\sigma\chi} + w_\chi^2 f_{NL}^\chi \,,
\ee
where we write the different contributions as
\ba
 \label{fNLsigma}
\dfrac{6}{5} f_{NL}^\sigma 
 &\equiv&  \dfrac{N_{\sigma \sigma}}{N_\sigma^2}
 = \dfrac{1}{f} \left[ \dfrac{3}{2}\left( 1 + \dfrac{g'' g}{g'^2}\right) -2f- f^2 \right] 
\\
 \label{fNLsigmachi}
\dfrac{6}{5} f_{NL}^{\sigma\chi}
 &\equiv& \dfrac{N_{\chi \sigma}}{N_\chi N_\sigma} 
 = \dfrac{(1-f)(3+f)}{f} 
\\
 \label{fNLchi}
\dfrac{6}{5} f_{NL}^\chi
 &\equiv& \dfrac{N_{\chi \chi}}{N_\chi^2}
 = \dfrac{1}{f} \left[ 9\left(1 - \frac23 \dfrac{\Gamma''\Gamma}{\Gamma'^2} \right) - 2f-f^2 \right] 
\ea
where we have used Eqs.~(\ref{Na}-\ref{Nchichi}).

In the limit $w_\sigma\to1$ then $f_{NL}\to f_{NL}^\sigma$ and we recover the standard result for the curvaton \cite{Bartolo:2003jx,Lyth:2005fi,Sasaki:2006kq}. In the opposite limit where $w_\chi\to1$ and $f\to1$ we recover the standard result for modulated reheating \cite{Zaldarriaga:2003my}
\be
f_{NL}^\chi \to 5 \left[ 1 - \dfrac{\Gamma''\Gamma}{\Gamma'^2}  \right] \,.
\ee

%%%%%%%%%%%%%%%%%%%%%%%%%%%%%%%%%%%%%%%%%%%%%%%%%%
\begin{figure}
\centering
\includegraphics[scale=.55]{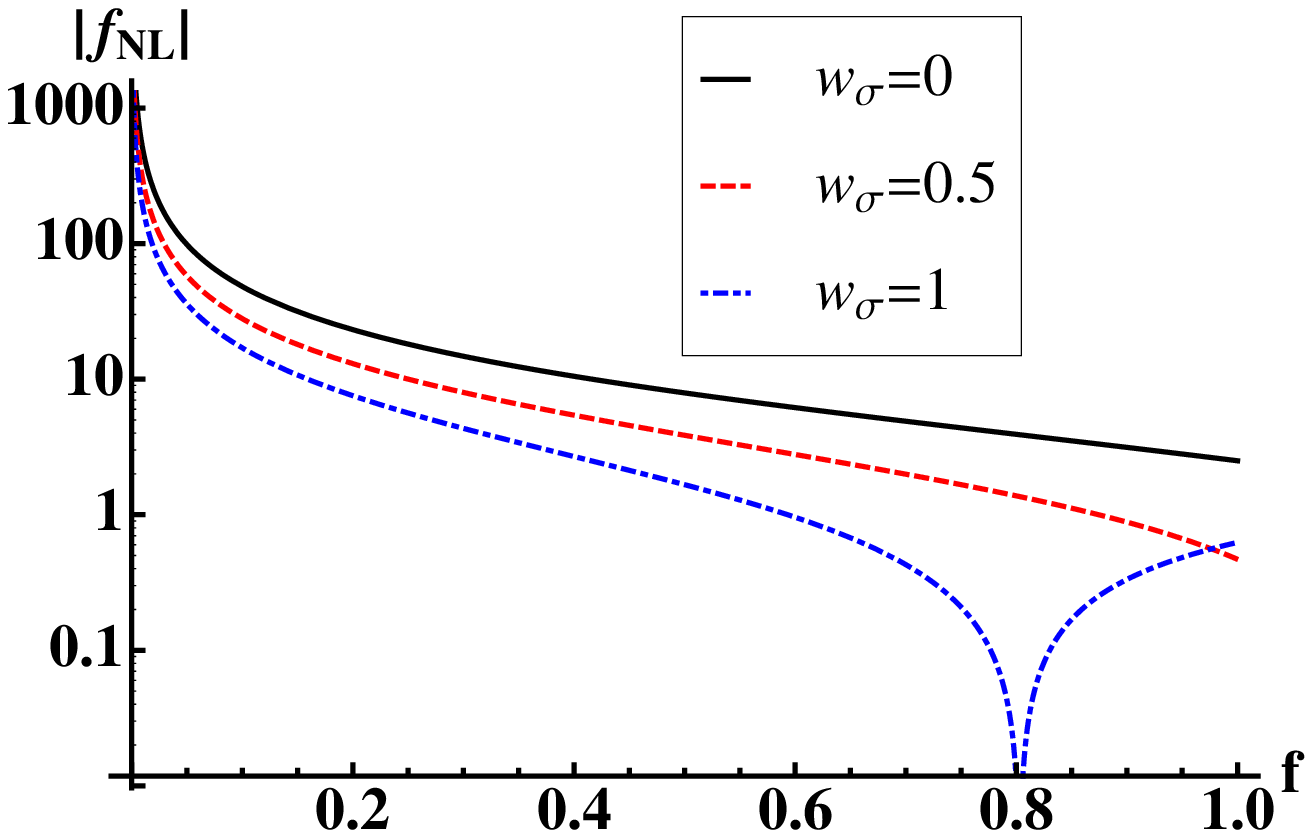}
\includegraphics[scale=.55]{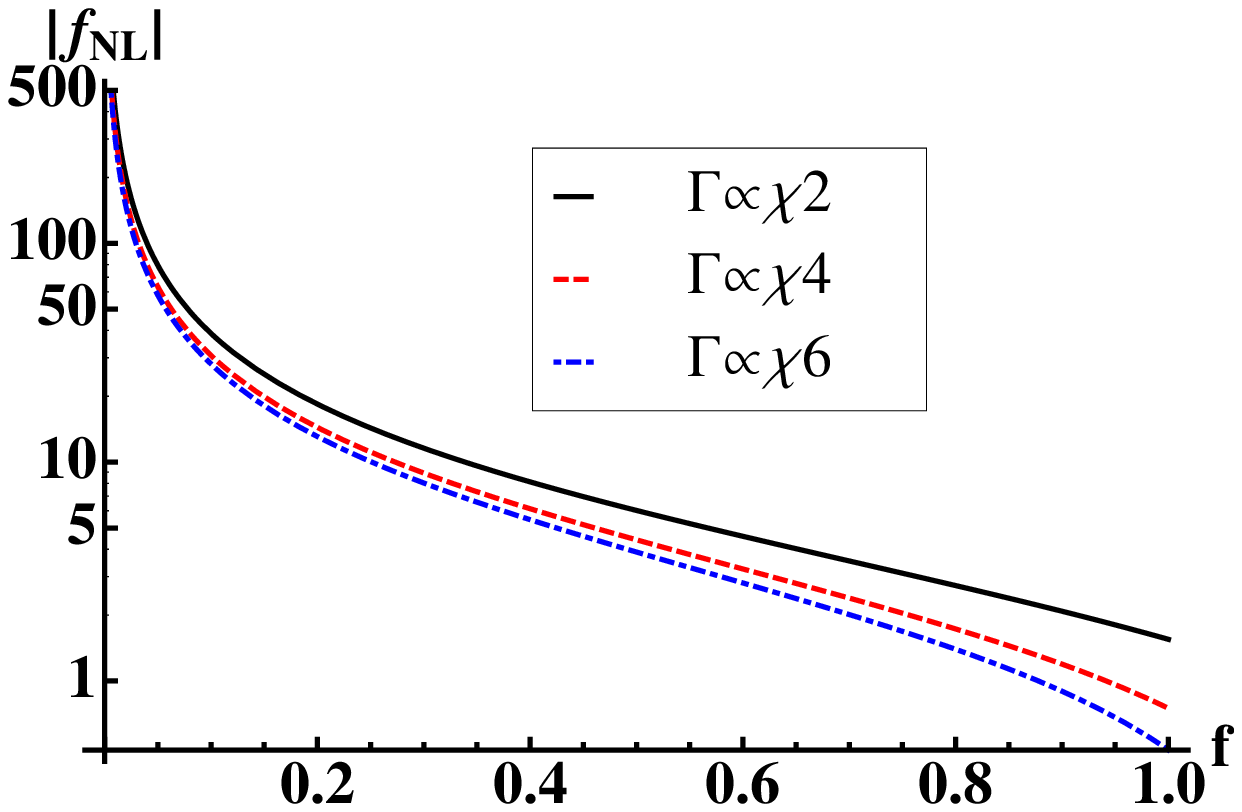}
\caption{A logarithmic plot for $f_{NL}$ as a function of relative energy density $f$. In both plots we assumed  $g \propto \sigma$  and neglected the contribution from inflaton field ($w_\phi\simeq 0$). For the left plot we also assumed $\Gamma \propto \chi^2$ and in the right one, we have set $w_\sigma=0.2$. The apparent singularity in the left plot is due to a change of  sign in $f_{NL}$.}
\label{fNLplot}
\end{figure}
%%%%%%%%%%%%%%%%%%%%%%%%%%%%%%%%%%%%%%%%%%%%%%%%%%

The primordial trispectrum is composed of distinct two terms \cite{Byrnes:2006vq}
\ba
\tau_{NL} = \frac{N_{ab}N^{ac}N_cN^b}{(N_a N^a)^3} \,,\\
g_{NL} = \frac{25}{54} \frac{N_{abc}N^aN^bN^c}{(N_a N^a)^3} \,.
\ea
which in our case we can write as
\ba
 \label{tauNL}
\frac{25}{36} \tau_{NL} &=& w_\sigma^3 (f_{NL}^\sigma)^2 +2w_\sigma^2 w_\chi f_{NL}^\sigma f_{NL}^{\sigma\chi}  \nonumber\\
&& + w_\sigma w_\chi ( w_\sigma + w_\chi ) (f_{NL}^{\sigma\chi} )^2 +
2 w_\sigma w_\chi^2 f_{NL}^{\sigma\chi} f_{NL}^\chi + w_\chi^3 (f_{NL}^\chi)^2 \,,\\
g_{NL} &=& w_\sigma^3 g_{NL}^\sigma +3w_\sigma^2 w_\chi g_{NL}^{\sigma\sigma\chi} +
3 w_\sigma w_\chi^2 g_{NL}^{\sigma\chi\chi} + w_\chi^3 g_{NL}^\chi \,,
\ea
where we identify the different contributions to $g_{NL}$ as
\ba
\label{gsig}
\frac{54}{25} g_{NL}^\sigma &\equiv& \frac{N_{\sigma\sigma\sigma}}{N_\sigma^3} 
 \,,\nonumber\\
 &=& \frac{9}{4f^2} \left( \frac{g^{\prime\prime\prime}g^2}{g^{\prime3}} +3 \frac{g^{\prime\prime}g}{g^{\prime2}} \right) - \frac{9}{f} \left( 1+\frac{g^{\prime\prime}g}{g^{\prime2}} \right) +\frac{1}{2} \left( 1-9\frac{g^{\prime\prime}g}{g^{\prime2}} \right) +10f + 3f^2 
\,,\\
\label{gsigsigchi}
\frac{54}{25} g_{NL}^{\sigma\sigma\chi} &\equiv& \frac{N_{\sigma\sigma\chi}}{N_\sigma^2N_\chi} 
 \,,\nonumber\\
 &=& \frac{3(1-f)(3+f)}{2f^2} \left( 1 + \frac{g^{\prime\prime}g}{g^{\prime2}} -\frac{8f}{3} -2f^2 \right)
\,,\\
\label{gsigchichi}
\frac{54}{25} g_{NL}^{\sigma\chi\chi} &\equiv& \frac{N_{\sigma\chi\chi}}{N_\sigma N_\chi^2} 
 \,,\nonumber\\
 &=& \frac{(1-f)(3+f)}{f^2} \left( 9 - 6 \frac{\Gamma^{\prime\prime}\Gamma}{\Gamma^{\prime2}}  -4f - 3f^2 \right) 
\,,\\
\label{gchi}
\frac{54}{25} g_{NL}^\chi &\equiv& \frac{N_{\chi\chi\chi}}{N_\chi^3} 
\,,\nonumber\\
 &=&  \frac{1}{f^2} \left\{ 
%(9-4f-3f^2)(1-f)(3+f) + 12(9-2f-f^2) 
135-54f-22f^2+10f^3+3f^4
% -6 \left[ (1-f)(3+f)+24-4f-2f^2 \right] 
-18 ( 9-2f-f^2 )
\frac{\Gamma^{\prime\prime}\Gamma}{\Gamma^{\prime2}}  + 36 \frac{\Gamma^{\prime\prime\prime}\Gamma^2}{\Gamma^{\prime3}} \right\}   
\,.
\ea
where we have used Eqs.~(\ref{Nsigsig}-\ref{fchi}).

%%%%%%%%%%%%%%%%%%%%%%%%%%%%%%%%%%%%%%%%%%%%%%%%%%
\begin{figure}
\centering
\includegraphics[scale=.5]{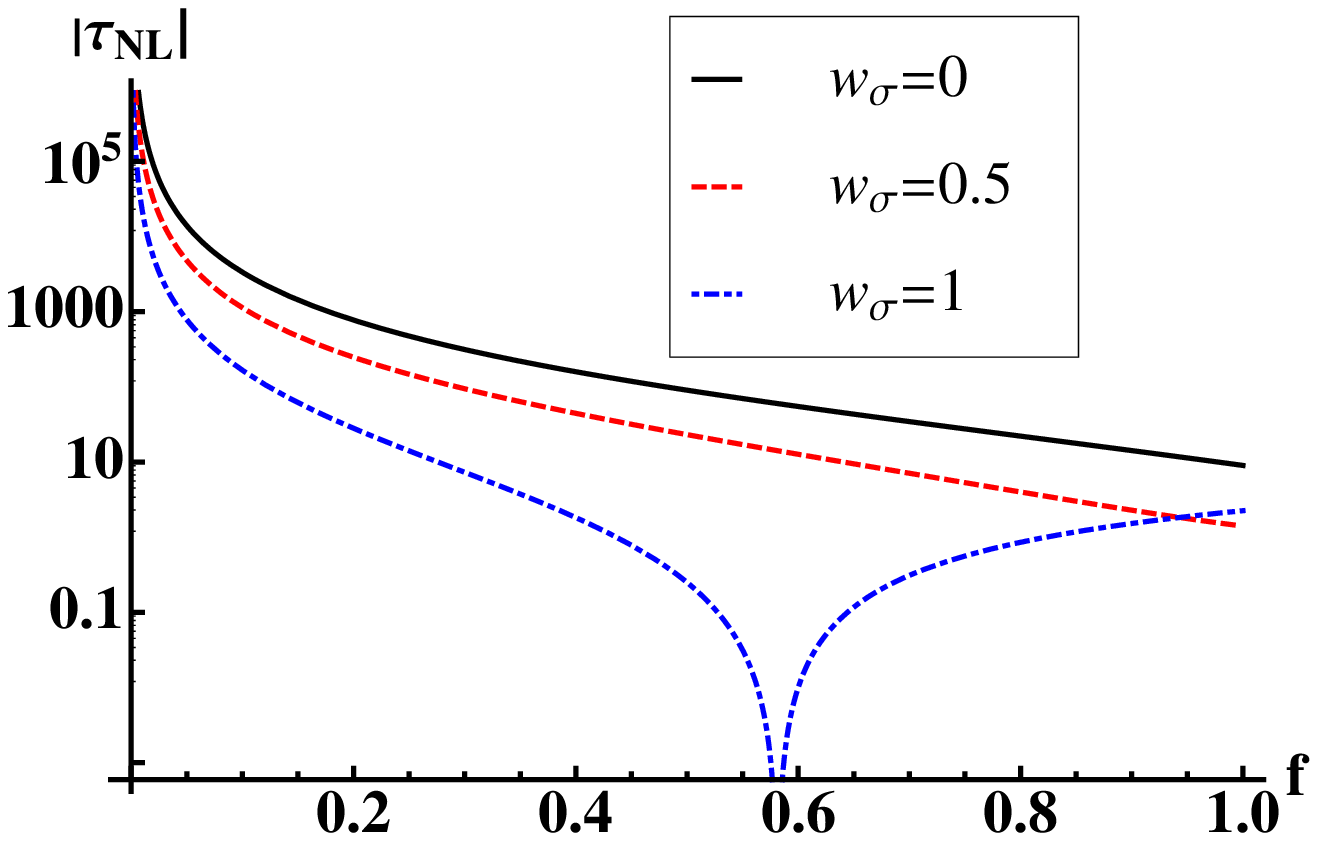}
\includegraphics[scale=.55]{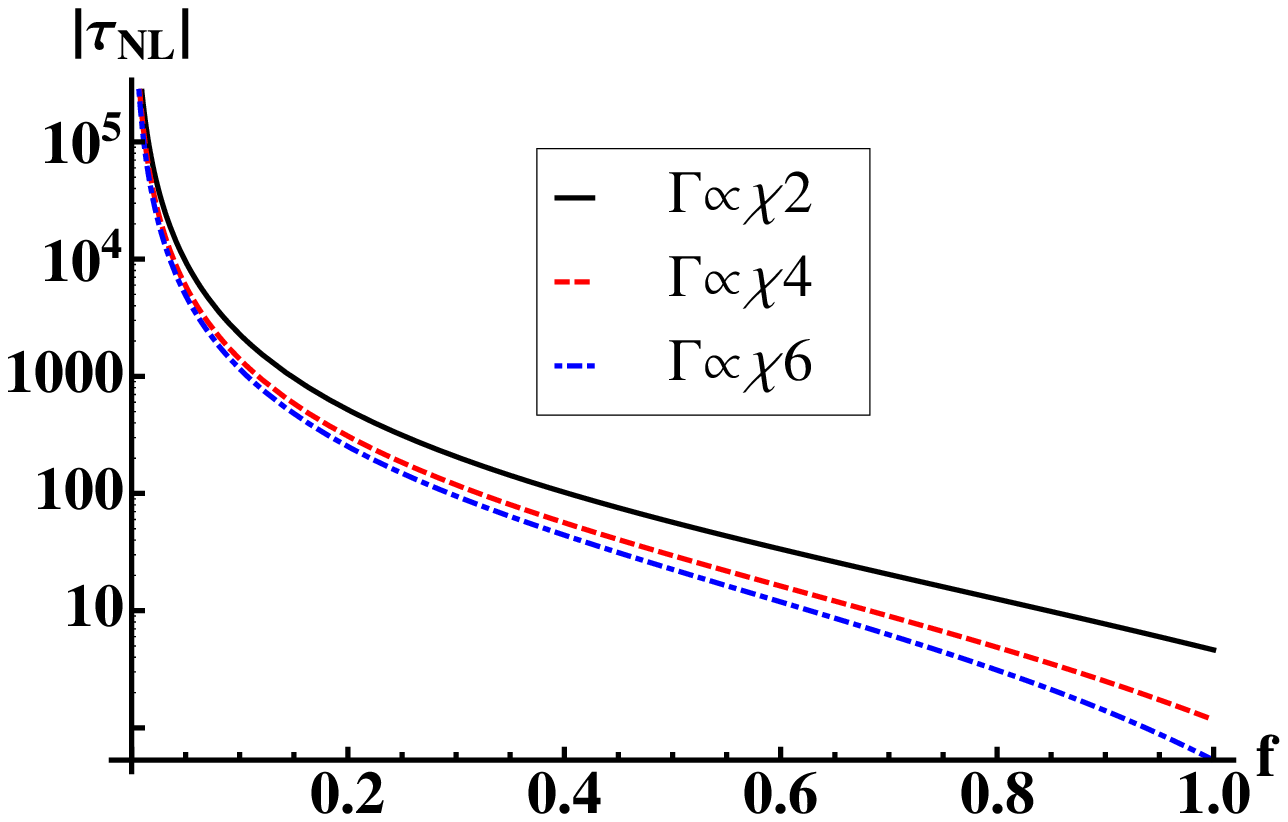}
\caption{A plot for $\tau_{NL}$ as a function of relative energy density $f$. The values of  the independent parameters are the same as in Fig.\ref{fNLplot}}
\end{figure}
%%%%%%%%%%%%%%%%%%%%%%%%%%%%%%%%%%%%%%%%%%%%%%%%%%
%%%%%%%%%%%%%%%%%%%%%%%%%%%%%%%%%%%%%%%%%%%%%%%%%%
\begin{figure}
\centering
\includegraphics[scale=.5]{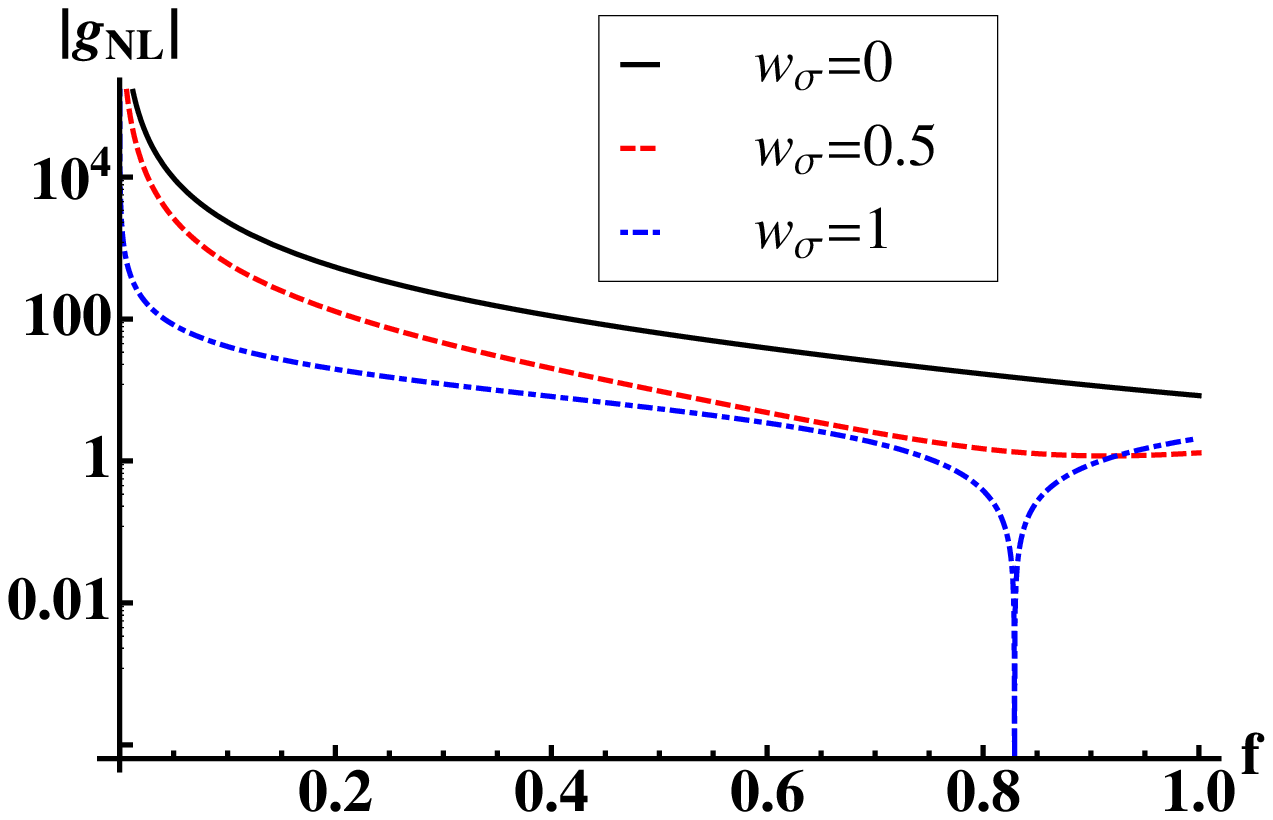}
\includegraphics[scale=.55]{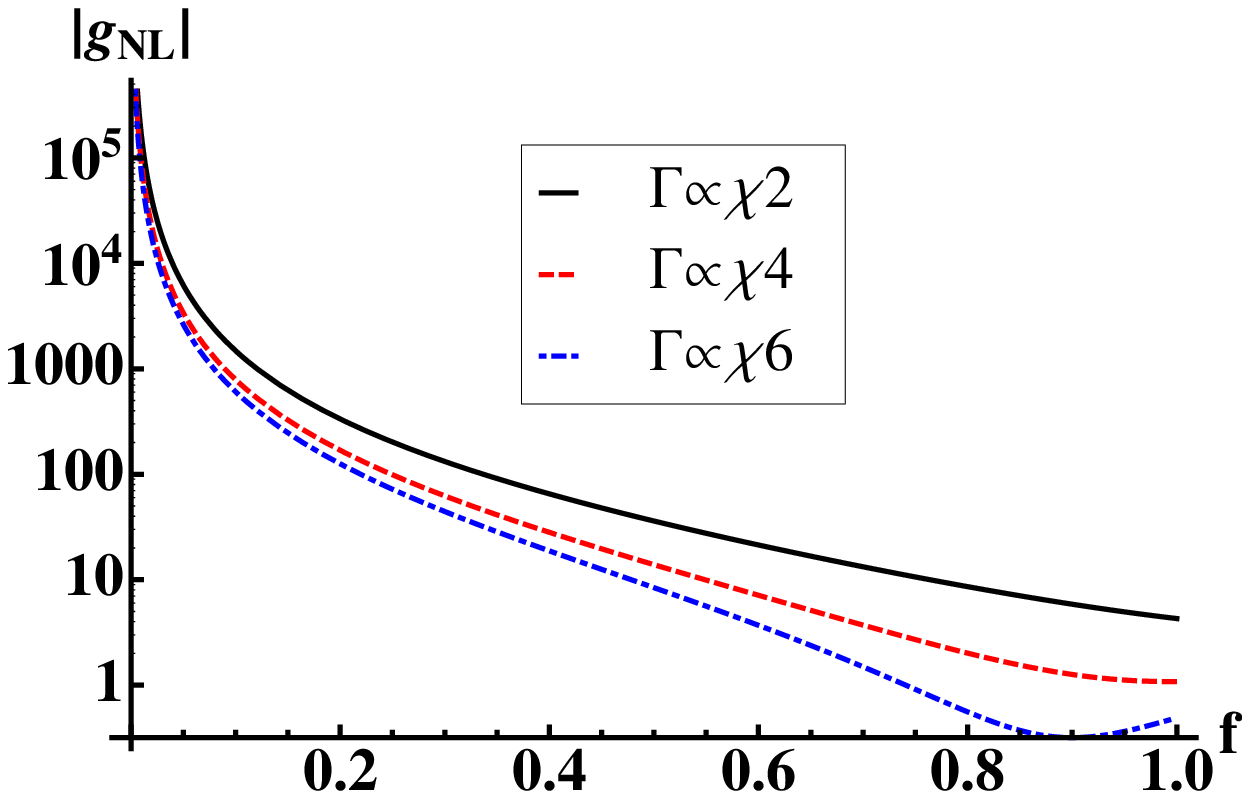}
\caption{A plot for $g_{NL}$ as a function of relative energy density $f$. Again the values of  the independent parameters are the same as in Fig.\ref{fNLplot} }
\end{figure}
%%%%%%%%%%%%%%%%%%%%%%%%%%%%%%%%%%%%%%%%%%%%%%%%%%

In the curvaton limit $w_\sigma\to1$ then $g_{NL}\to g_{NL}^\sigma$ and we recover the standard result for the curvaton \cite{Sasaki:2006kq}. 
In the opposite limit, $w_\chi\to1$, we recover the result for modulated reheating when $f\to1$ \cite{Suyama:2007bg,Wands:2010af}
\be
g_{NL}^\chi \to \frac{50}{3} \left[ 2 - 3\dfrac{\Gamma^{\prime\prime}\Gamma}
{\Gamma'^2}   +\dfrac{\Gamma^{\prime\prime\prime}\Gamma^2}{\Gamma'^3} \right] \,.
\ee

%%%%%%%%%%%%%%%%%%%%%%%%%%%%%%%%%%%%%%%%%%%%%%%%%%
\subsection{The Suyama-Yamaguchi inequality}

Here we verify that the Suyama-Yamaguchi (SY) inequality \cite{Suyama:2007bg},  
stating that $\tau_{NL} \ge (\frac{6 }{5} f_{NL})^2$,  holds at tree-level in our model~\footnote{Note that loop-corrections violate this tree-level inequality \cite{Byrnes:2011ri,Tasinato:2012js}.}.

A direct analysis shows that
\ba
\label{SY}
K\equiv \frac{25}{36} \tau_{NL} - f_{NL}^2 = c_1 ({f_{NL}^{\chi }})^2  + c_2 f_{NL}^{\chi } + c_3 
\ea
in which
\ba
\label{c1}
c_1  &&\equiv w_\chi^3 (1- w_\chi)  \\
c_2  &&\equiv  2w_\sigma w_\chi^2 \left[    (1- 2 w_\chi) f_{NL}^{\sigma \chi}  - w_\sigma 
f_{NL}^\sigma  \right]  \\
c_3  &&\equiv  w_\sigma^3 (1- w_\sigma) {f_{NL}^{\sigma \, 2 }} + w_\sigma w_\chi (w_\sigma + w_\chi - 4 w_\sigma w_\chi) {f_{NL}^{\sigma \chi  \, 2}}
+ 2 w_\chi w_\sigma^2  (1- 2 w_\sigma)   f_{NL}^{\sigma}    f_{NL}^{\sigma \chi}
\ea
We wish to determine the sign of $K$ in Eq.~(\ref{SY}) which is a quadratic function 
of $f_{NL}^{\chi }$ for $c_1\neq0$.  Hence we re-write Eq.~(\ref{SY}) as
\be
\label{Ksquare}
K = c_1 \left( f_{NL}^\chi+\frac{c_2}{2c_1}\right)^2 + \frac{\Delta}{c_1} \,,
\ee
where
\ba
\label{disc}
\Delta = w_\chi^3 w_\sigma (1-w_\sigma - w_\chi) ( w_\sigma f_{NL}^{\sigma} + w_\chi f_{NL}^{\sigma\chi})^2  \,.
\ea
Both the coefficient, $c_1$ in Eq.~(\ref{c1}), and the discriminator, $\Delta$ in Eq.~(\ref{disc}) are non-negative, since the weights, $w_\sigma$ and $w_\chi$, and their sum, $w_\sigma+w_\chi$, are bounded between zero and one. Hence $K$ given in Eq.~(\ref{Ksquare}) is non-negative and we conclude that  $\tau_{NL} \ge  (\frac{6 }{5} f_{NL})^2$ as required. 

One may ask under what conditions the SY inequality is saturated. Firstly this can occur if $c_1=0$ and $c_2 f_{NL}^{\chi } + c_3=0$, which requires either $w_\sigma=1$ or $w_\chi=1$, i.e., either the curvaton fluctuations or the modulator fluctuations dominate the primordial density perturbations corresponding to effectively a single source for the primordial density field, or $w_\phi=1$ corresponding to the Gaussian case,  $\tau_{NL}=f_{NL}=0$

For $c_1\neq0$, the SY equality is saturated when $\Delta =0$ and $2c_1f_{NL}^\chi+c_2=0$ in Eq.~(\ref{Ksquare}). This either requires
\be
w_\sigma^2 f_{NL}^{\sigma} = w_\chi^2 f_{NL}^\chi = - w_\sigma w_\chi f_{NL}^{\sigma\chi}
\ee
which implies $\tau_{NL}=f_{NL}=0$, or requires $w_\sigma+w_\chi=1$ and
\ba
\label{const}
w_\sigma f_{NL}^{\sigma} - w_\chi f_{NL}^{\chi}= (w_\sigma - w_\chi) f_{NL}^{\sigma \chi}
\ea
or, equivalently,
\be
\label{SY-w}
w_\sigma = \frac{f_{NL}^\chi-f_{NL}^{\sigma\chi}}{f_{NL}^\sigma+f_{NL}^\chi-2f_{NL}^{\sigma\chi}} \,, \quad 
w_\chi = \frac{f_{NL}^\sigma-f_{NL}^{\sigma\chi}}{f_{NL}^\sigma+f_{NL}^\chi-2f_{NL}^{\sigma\chi}} \,.
\ee
Note that this is possible only when $f_{NL}^\sigma$ and $f_{NL}^\chi$ are either both less than $f_{NL}^{\sigma\chi}$ or both greater than $f_{NL}^{\sigma\chi}$,
i.e., 
\be
\left( f_{NL}^\sigma-f_{NL}^{\sigma\chi} \right) \left( f_{NL}^\chi-f_{NL}^{\sigma\chi} \right) > 0 \,.
\ee

%Now the condition $\Delta =0$ can happen either 
%$w_\phi=0$ or $w_\sigma f_{NL}^{\sigma} - w_\chi f_{NL}^{\sigma} =0$. Therefore, if $w_\phi \neq 0$, in order to saturate the SY inequality we require $w_\sigma f_{NL}^{\sigma} - w_\chi f_{NL}^{\chi}=0$. Combined with Eq. (\ref{const}) this can be consistent  either $1 - 2 w_\chi=0$  or $ f_{NL}^{\sigma \chi}=0$. On the other hand,  consider the particular case in which $w_\phi=0$ so $w_\sigma + w_\chi =1$. Then the SY inequality is saturated only if Eq. (\ref{const}) is met.  In Fig. \ref{SYfig} we have verified the saturation of the SY inequality bound when the condition (\ref{const}) is met when $w_\phi=0$ for arbitrary $w_\sigma$. 

%%%%%%%%%%%%%%%%%%%%%%%%%%%%%%%%%%%%%%%%%%%%%%%%%%

\section{Discussion}
\label{sect:discuss}

The curvaton and modulated reheating scenarios have previously been studied as distinct models for the origin of structure from quantum field fluctuations during inflation. Here we have considered a curvaton scenario where the curvaton decay rate may be modulated by a second scalar field. Thus fluctuations in two independent fields may be responsible for both the primordial power spectrum and primordial non-Gaussianity described by higher-order correlations. 

The relative contributions to the primordial power spectrum \eqref{Pzeta} are determined by the weights, $w_\sigma$ and $w_\chi$ defined in \eqref{wa}. 
The overall contribution of both curvaton and modulator fluctuations to the first-order primordial density perturbation is proportional to the fractional density in the curvaton at the time of decay, $f$ defined in \eqref{fdef}. The relative contribution depends on the fractional perturbations in the curvaton amplitude of oscillation, $g$, versus the decay rate, $\Gamma$:
\be
\frac{w_\sigma}{w_\chi} = \frac{g'/g}{\Gamma'/\Gamma}\,.
\ee
We recover (i) previous results for the curvaton scenario in the limit that curvaton field fluctuations dominate the primordial power spectrum ($w_\sigma\gg w_\chi$) and (ii) previous results for modulated reheating in the limit that the curvaton dominates the total energy when it decays ($f\to1$) {\em and} modulated fluctuations dominate ($w_\chi\gg w_\sigma$) .

We also allow for an adiabatic density perturbation  produced due to inflaton perturbations during inflation, $\zeta_\phi$. This is required to be Gaussian, hence the inflaton contributes to the first-order density perturbation, but not to higher-order correlators. Any detection of local-type primordial non-Gaussianity would be evidence of non-adiabatic field fluctuations playing a role in the origin of large-scale structure, see however \cite{Namjoo:2012aa}.

Non-Gaussianity can become very large, either for $f\ll 1$ or due to non-linear evolution, $g''g\gg g^{\prime2}$, in the curvaton scenario. Indeed it is bounded by current  observations \cite{Bennett:2012fp}
\be
f_{NL} =  37.2\pm19.9 \,.
\ee
Such a large value is difficult to achieve in standard modulated reheating unless the decay rate is strongly dependent upon the modulator field, $\Gamma^{\prime\prime}\Gamma/\Gamma^{\prime2}\sim 8$. On the other hand one can easily have large non-Gaussianity in a modulated curvaton reheating when $f\ll1$ even when the modulator dominates the primordial fluctuations, $w_\chi\to1$.

In the simplest scenario of a quadratic curvaton potential, leading to linear evolution of the curvaton field, $g''=0$, and a linear coupling, $\lambda(\chi)$ in Eq.~(\ref{L}), leading to a quadratic decay rate, $\Gamma\propto \chi^2$, the primordial non-Gaussianity is a function of $f$, $w_\sigma$ and $w_\chi$. Hence a measurement of the three lowest-order non-Gaussian correlators, $f_{NL}$, $\tau_{NL}$ and $g_{NL}$, would be required to determine $f$, $w_\sigma$ and $w_\chi$. 

There are some general predictions for the simplest model of a quadratic curvaton potential with $g''=0$. The contributions to the primordial bispectrum, given in Eq.~\eqref{fNL}, from $f_{NL}^\chi$ and $f_{NL}^{\sigma\chi}$, defined in Eqs.~(\ref{fNLchi}) and (\ref{fNLsigmachi}), are non-negative for any $\Gamma''\Gamma\leq\Gamma^{\prime2}$, which includes a quadratic decay rate. Thus in the modulated curvaton decay when $g''=0$, we find a lower bound $f_{NL}\geq f_{NL}^\sigma\geq -5/4$, generalising the result found previously for a single curvaton \cite{Sasaki:2006kq}, and multiple curvaton decays~\cite{Assadullahi:2007uw}. This bound is saturated, $f_{NL}=-5/4$, for $w_\sigma=1$ and $f=1$.

For this simple quadratic curvaton scenario with linear evolution, such that $g''$ and $g'''$ can be neglected, we find the third-order trispectrum parameter $g_{NL}^\sigma\propto f^{-1} \propto f_{NL}$ whereas the second-order trispectrum parameter $\tau_{NL}\propto f^{-2}\propto f_{NL}^2$. Hence $\tau_{NL}$ is much larger than $g_{NL}^\sigma$ if $f_{NL}$ is large. On the other hand $g_{NL}^{\sigma\sigma\chi}\propto f^{-2}\propto f_{NL}^2$ even for $g''\simeq0$, and hence $g_{NL}/\tau_{NL}$ is not necessarily suppressed for a simple quadratic curvaton in a modulated curvaton scenario with $\delta\Gamma\neq0$.

We generally expect $g_{NL}\propto \tau_{NL} \propto f_{NL}^2$ due to non-linear evolution of the curvaton field with a self-interaction potential \cite{Enqvist:2008gk,Enqvist:2009zf,Enqvist:2009ww,Suyama:2010uj,Fonseca:2011aa}. On the other hand the self-interacting curvaton can give rise to strongly scale-dependent non-Gaussianity \cite{Byrnes:2011gh}, while the modulated curvaton decay with a quadratic curvaton potential gives rise to non-Gaussianity which is scale-independent.

We have verified that the trispectrum parameter $\tau_{NL}$ obeys the Suyama-Yamaguchi inequality \cite{Suyama:2007bg} $\tau_{NL}\geq (36/25)f_{NL}^2$. It is saturated when the curvaton perturbations ($w_\sigma\to1$) or modulator perturbations ($w_\chi\to1$) dominate, or in the trivial case of Gaussian perturbations when $w_\phi\to1$. It can also be satisfied for particular parameter values even when both curvaton and modulator fluctuations contribute to the primordial density field, $w_\sigma+w_\chi=1$ given by Eq.~(\ref{SY-w}). This emphasizes that a single-source for the primordial density field is a sufficient condition for the SY inequality to be saturated, but not a necessary condition \cite{Suyama:2007bg}.

%%%%%%%%%%%%%%%%%%%%%%%%%%%%%%%%%%%%%%%%%%%%%%%%%%
\begin{figure}
\label{SYfig}
%\centering
\includegraphics[scale=.6]{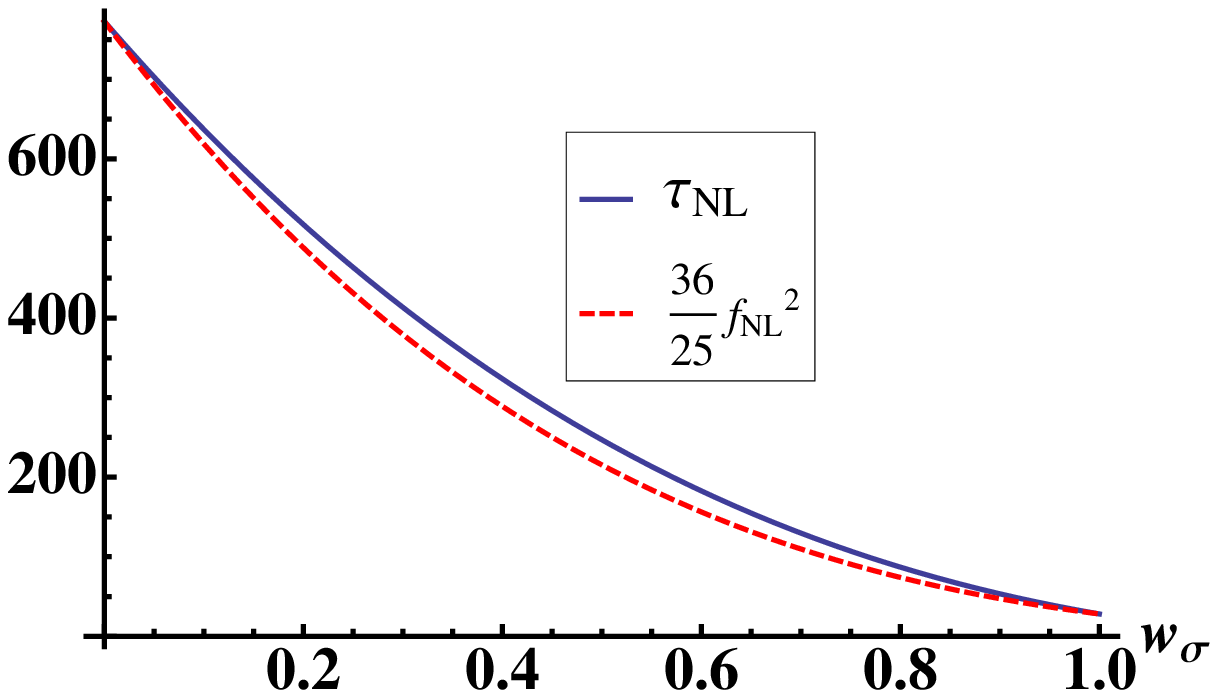}
\includegraphics[scale=.5]{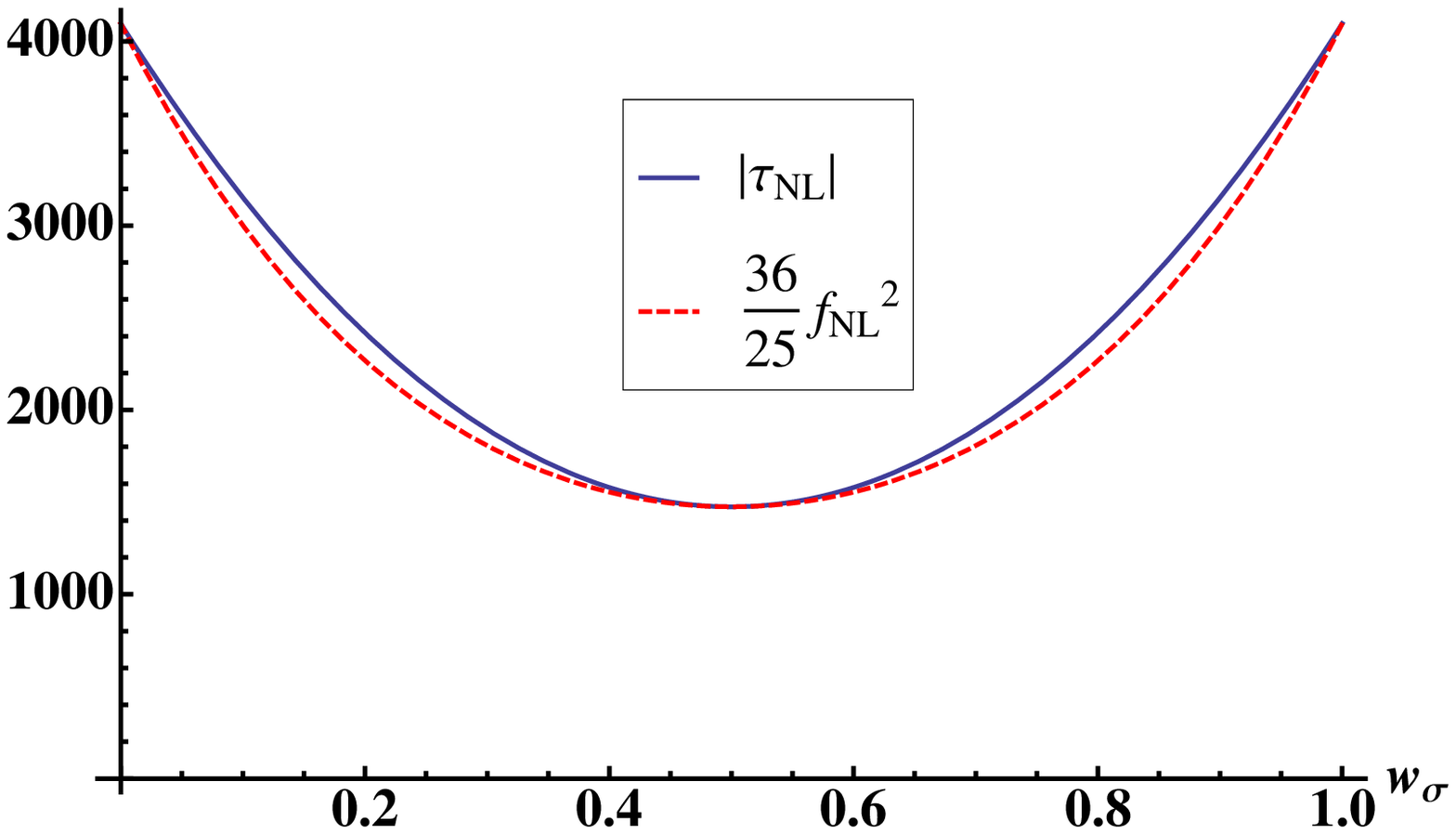}
\caption{Plots for the  Suyama-Yamaguchi equality/inequality (\ref{SY}). In both figures $w_\phi=0$. Left:  $f=0.2$ with generic value of $f_{NL}^\sigma, f_{NL}^\chi$ and 
$f_{NL}^{\sigma \chi}$.  Right: The particular case $f_{NL}^\sigma = f_{NL}^\chi = 5 f_{NL}^{\sigma\chi}$  is plotted for arbitrary $w_\sigma$.  As expected the SY inequality is saturated when $w_\sigma =0, 1$ or when  Eq. (\ref{const}) is met 
with $w_\sigma + w_\chi=1$, which in this particular case, yields $w_\sigma = w_\chi =0.5$. }
\end{figure}
%%%%%%%%%%%%%%%%%%%%%%%%%%%%%%%%%%%%%%%%%%%%%%%%%%

Throughout this work we have assumed that the decay products of the curvaton rapidly thermalise leaving no residual isocurvature perturbations. If the matter asymmetry inherits a different density perturbation from the overall radiation density (because it comes exclusively from either the curvaton decay products or the pre-existing radiation before the curvaton decay, but not both) then it may leave a residual matter isocurvature perturbation \cite{Lyth:2002my,Langlois:2008vk,Langlois:2012tm}. This would be further evidence of the origin of structure from non-adiabatic field perturbations during inflation, and measurements of the relative amplitude of residual isocurvature perturbations and their correlations with the adiabatic density perturbation could give independent constraints on model parameters. It would be interesting to investigate whether these could then distinguish modulated curvaton decay from standard curvaton or modulated reheating scenarios.

Although we have assumed that the modulator field has negligible energy density, it would be interesting to consider possible observational signatures of the eventual decay of the modulator field, assuming that it too eventually decays into standard model particles \cite{Enomoto:2012uy}, analogous to our recent study of the effect of the late-time decay of a field responsible for the inhomogeneous end of hybrid inflation \cite{Assadullahi:2012yi}. We leave this for future work.

{\em Note added: } While completing this work, we became aware of related work by Langlois and Takahashi \cite{Langlois2013}. Both papers should appear on the arXiv on the same day.

\acknowledgments

% We would like to thank XXX  for useful discussions. 
%
HF would like to thank the ICG, Portsmouth, for hospitality while this work was initiated and later finalized. 
HF, MHN and DW are grateful to the organisers of the YITP long-term workshop on Gravitation and Cosmology 2012, YITP-T-12-03.
MHN is in part supported by Yukawa Institute for Theoretical Physics (YITP), Kyoto University, under the Exchange
Program for Young Researchers of YITP.
DW is supported by STFC grant ST/H002774/1.

\end{document}